\documentclass{optica-article}

\journal{opticajournal} 

\articletype{Research Article}

\usepackage{lineno}
\usepackage{hyperref}
\hypersetup{
    colorlinks=true,         
    linkcolor=blue,          
    filecolor=blue,       
    urlcolor=blue,           
    citecolor=blue            
}

\begin{document}



\title{Modifying electronic and structural properties of 2D van der Waals materials via cavity quantum vacuum fluctuations: A first-principles QEDFT study}

\author{Hang Liu,\authormark{1,5*} Simone Latini,\authormark{1,2} I-Te Lu,\authormark{1} Dongbin Shin,\authormark{1,3,$\dag$} and Angel Rubio\authormark{1,4,$\ddag$}}

\address{\authormark{1}Max Planck Institute for the Structure and Dynamics of Matter and Center for Free-Electron Laser Science, Luruper Chaussee 149, 22761, Hamburg, Germany\\
\authormark{2}Department of Physics, Technical University of Denmark, 2800 Kgs. Lyngby, Denmark\\
\authormark{3}Department of Physics and Photon Science, Gwangju Institute of Science and Technology (GIST), Gwangju 61005, Republic of Korea\\
\authormark{4}Initiative for Computational Catalysis (ICC), The Flatiron Institute, 162 Fifth avenue, New York NY 10010, USA \\
\authormark{5}Songshan Lake Materials Laboratory, Dongguan, Guangdong 523808, China
}

\email{\authormark{*}hang.liu@mpsd.mpg.de\\
\authormark{$\dag$}dshin@gist.ac.kr\\
\authormark{$\ddag$}angel.rubio@mpsd.mpg.de} 

\begin{abstract*} 
Structuring the photon density of states and light-matter coupling in optical cavities has emerged as a promising approach to modifying the equilibrium properties of materials through strong light-matter interactions.
In this article, we employ state-of-the-art quantum electrodynamical density functional theory (QEDFT) to study the modifications of the electronic and structural properties of two-dimensional (2D) van der Waals (vdW) layered materials by the cavity vacuum field fluctuations.  
We find that cavity photons modify the electronic density through localization along the photon polarization directions, a universal effect observed for all the 2D materials studied here. 
This modification of the electronic structure tunes the material properties, such as the shifting of energy valleys in monolayer h-BN and 2H-MoS$_2$, enabling tunable band gaps. 
Also, it tunes the interlayer spacing in bilayer 2H-MoS$_2$ and T$_\text{d}$-MoTe$_2$, allowing for adjustable ferroelectric, nonlinear Hall effect, and optical properties, as a function of light-matter coupling strength. 
Our findings open an avenue for engineering a broad range of 2D layered quantum materials by tuning vdW interactions through fluctuating cavity photon fields.
\end{abstract*}

\section{Introduction}
Strong light-matter interactions can modify material properties through various microscopic mechanisms \cite{Bloch2022_Nature}. 
Illuminating a condensed matter system with an external ultrafast and ultrastrong light pulse can dynamically drive phase transitions in non-equilibrium states, where extensive studies have shown that electronic, magnetic, and superconducting phases can be modified and controlled through the coupling of light with phonons and electrons \cite{Bloch2022_Nature,Andre2011,Andre2022,Basov2017,Andrea2021,Basov2021}.
For instance, terahertz light pumping has been shown to induce shear distortions, leading to the formation of Weyl nodal points from a topologically trivial phase in the semimetal T$_\text{d}$-WTe$_2$ \cite{Sie2019_WTe2,Guan_WTe2_light_Weyl}.
Also, experiments have observed transient light-induced metastable magnetization in FePS$_3$ \cite{Emil_Nature}, a transition between quantum paraelectric and ferroelectric phases in SrTiO$_3$ \cite{Xian2019_ferro,Nova2019_ferro,PhysRevLett.129.167401}, and the light-induced superconductivity in K$_3$C$_{60}$ \cite{mitrano2016possible,Cavalleri02012018}. 
While light-driven processes provide a means of manipulating material properties, the phenomena induced by classical light fields are typically short-lived and exist in excited nonequilibrium states \cite{de2021colloquium,bao2022light,nuske2020floquet,galler2023mapping}.

Fluctuating photon fields within an optical cavity are expected to statically modify properties of a material in its ground state \cite{Bloch2022_Nature,RN9055,Hannes_NM_comment,hubener2024quantum,lu2025cavityengineeringsolidstatematerials}.
It has been demonstrated that the coupling between Landau levels and fluctuating cavity photons can disrupt both integral and fractional quantization of Hall states in electron gases \cite{Faist_cavity_QHE,RN9831}. 
Recent experiments also reveal that embedding bulk 1T-TaS$_2$ crystals in an optical cavity can effectively tune the transition temperature between conducting and insulating charge density wave phases, without exciting any degrees of freedom in the material \cite{Jarc2023_TaS2_cavity,TaS2_CDW}. 
Meanwhile, theoretical studies suggest various material properties mediated by fluctuating vacuum electromagnetic fields  \cite{Sentef}.
Cavity-induced effective spin-orbit coupling has been predicted to generate topologically massive Dirac cones in graphene \cite{graphene_Russian,graphene_Sentef,graphene_cavity_topology_single_particle,graphene_Vasil}, while strong photon-phonon interactions can drive a cavity-induced para-ferroelectric transition in SrTiO$_3$ by localizing lattice wavefunctions \cite{Simone_PNAS,PhysRevX.10.041027}, in a manner analogous to dynamical localization \cite{dunlap1986dynamic}. 
Cavities can also be used to control the magnetic state of the proximate quantum spin liquid $\alpha\text{-RuCl}_3$ \cite{vinas2023controlling} and to modify the quantum criticality of a honeycomb bilayer antiferromagnet \cite{weber2023cavity}. 
Notably, the recently developed quantum electrodynamical density functional theory (QEDFT) enables first-principles investigations of strong light-matter interactions in realistic materials within optical cavities or other electromagnetic environments \cite{QED_functional_Christian,ab-initio_Perspective_Dominik,I-Te_PRA_QEDFT}. 
Using this \textit{ab-initio} theoretical modeling method, the enhancement of superconductivity in MgB$_2$ from photon-mediated electron-phonon coupling was investigated \cite{I-Te_PNAS}. 
These experimental observations and theoretical predictions indicate that optical cavities can indeed statically modify material properties by coupling vacuum photon field fluctuations and matter.
Especially, as a large category of quantum materials with intriguing properties \cite{2dmaterials}, the 2D layered materials, stacked by the van der Waals forces, remain unexplored in terms of non-perturbative phenomena induced by the coupling with cavity fluctuating photon fields.

In this article, we employ state-of-the-art QEDFT simulations to explore how the fluctuating cavity photons, arising from the quantum confinement in optical cavities, modify the electronic and structural properties of a large class of two-dimensional (2D) van der Waals (vdW) layered quantum materials, as well as one-dimensional (1D) chain. 
We first demonstrate that cavity-induced charge localization occurs in a 1D hydrogen chain, resulting in a narrowing of the energy width of electronic band. 
This effect can be extended to 2D layered material systems, where charge localization aligns with the polarization directions of cavity photon modes. 
We show that both electronic and atomic structures of monolayer h-BN and 2H-MoS$_2$ can be tuned, including band shifts and band gap modifications. Additionally, we investigate the changes in optical and ferroelectric properties in bilayer 2H-MoS$_2$ and bilayer T$_\text{d}$-MoTe$_2$, enabled by  cavity-controlled interlayer spacing through the tuned vdW interactions by structured cavity vacuum fields.
Unlike previous QEDFT studies that have primarily focused on molecules and 3D materials \cite{QED_functional_Christian,ab-initio_Perspective_Dominik,I-Te_PRA_QEDFT,I-Te_PNAS}, this work reveals how cavity-induced modifications affect 2D layered quantum materials, showing that van der Waals interactions can be influenced by electromagnetic cavity fluctuations.
Also, this study suggests that photon-induced dynamical charge localization offers a mechanism for statically controlling properties of 2D materials through strong light-matter coupling.

\section{Computational methods}
The QEDFT simulations are performed using our in-house modified Quantum Espresso package \cite{Giannozzi2017} with Perdew-Burke-Ernzerhof functional for electron-electron exchange interaction~\cite{Perdew1996}, where the simulations can also be carried out using the Octopus package \cite{tancogne2020octopus}. Electrons interact with photons via the minimal coupling $\hat{\mathbf{p}}_i-\hat{\mathbf{A}}$ with electron momentum operator $\hat{\mathbf{p}}_i=-i\nabla_{\mathbf{r}_i}$ and photon vector-potential operator $\hat{\mathbf{A}} = c \sum_\beta \frac{\lambda_\beta}{\sqrt{2\omega_\beta}} (\hat{a}_\beta^\dagger + \hat{a}_\beta) \mathbf{e}_\beta$ in Hartree atomic units, where $c$ is velocity of light, $\hat{a}_\beta$ ($\hat{a}^\dagger_\beta$) is photon annihilation (creation) operator, and $\lambda_\beta$, $\omega_\beta$, and $\mathbf{e}_\beta$ are mode strength, frequency, and polarization vector for the $\beta$th photon mode, respectively. The mode strength can be expressed by an effective mode volume via $\lambda_\beta = \sqrt{\frac{4 \pi}{V}}$ \cite{Christian_PNAS_QEDFT}. The electron-photon exchange functional under the local density approximation (LDA) is employed to account for quantum light-matter interactions, where the functional depends on the dimensionless ratio of the mode strength $\lambda_\beta$ and frequency $\omega_\beta$ \cite{Christian_PNAS_QEDFT,I-Te_PRA_QEDFT,I-Te_PNAS} for the cases of one photon mode and two photon modes with orthogonal polarization directions in this study. This implies that QEDFT yields the same outcomes for different values of $\lambda_\beta$ and $\omega_\beta$ as long as their ratio remains the same, thus the ratio $\frac{\lambda_\beta}{\omega_\beta}$ is used to characterize light-matter coupling strength in this work.
The number of electrons in one unit cell is used for the electron-photon exchange functional.
The Grimme-D3 type of vdW parametrization is adopted to correct the forces between layers of 2D crystalline materials \cite{Grimme_vdw}, while the LDA electron-photon exchange functional does not contain the photon-induced corrections on vdW parametrization.
Lattice constants of the 2D materials  outside cavities are relaxed until the cell pressure converges below 0.5 kbar; these lattice constants are used for modeling the materials inside cavities.  
The slab models of the 2D crystals have a vacuum layer greater than 15 \AA~to avoid spurious interactions between adjacent images.
Atomic structures are relaxed with a force convergence threshold of $6.0 \times 10^{-6}$ Ry/Bohr.
The plane-wave basis set has a kinetic energy cutoff of 80 Ry, and the first Brillouin zone in reciprocal space is sampled using a $6 \times 6 \times 1$ ($24 \times 24 \times 1$) $\mathbf{k}$-point mesh for structure relaxation and self-consistent field (non-self-consistent field) calculations. 
The energy convergence criterion for self-consistent iterations of electronic states is set to $2.0 \times 10^{-14}$ Ry. 
Within the employed QEDFT approach, different cavity configurations can be realized by changing the light-matter coupling strength $\frac{\lambda_\beta}{\omega_\beta}$ and the photon polarization vector $\mathbf{e}_\beta$ \cite{Christian_PNAS_QEDFT,I-Te_PRA_QEDFT,I-Te_PNAS}, where photon energy of $0.1$ eV is used for fluctuating cavity vacuum fields in this work.
Two representative types of cavity photon modes are adopted to interact with 2D materials: one mode with polarization along the out-of-plane $z$ direction, and two modes with polarizations along the in-plane $x$ and $y$ directions. $x$ and $y$ modes have the same strength and frequency. The symbol $\beta$ for mode index is omitted in the following for brevity.
The electron-photon exchange potential has a weight of 1 for each photon mode. 

To study cavity-modified properties of 2D materials, we further compute the optical absorption spectrum, Berry curvature dipole, injection current, and electric dipole moment. The absorbance 
$$\alpha_{\parallel} = \frac{\text{Re} \sigma_{\parallel}}{|1+\frac{\sigma_{\parallel}}{2}|^2}$$
is computed from the normalized optical conductivity $\sigma_{\parallel}=\frac{i\omega_p L}{c} (1-\varepsilon_{\parallel})$ \cite{matthes2016influence}, where $\omega_p$ is the frequency of the absorbed light field polarized parallel to 2D material plane, $c$ is the speed of light in vacuum, and $L$ is the slab thickness of simulation cell. The dielectric function $\varepsilon_{\parallel}$, computed from random-phase approximation using the code of epsilon.x from Quantum Espresso package, depends on frequency $\omega_p$. Electron-hole interaction is not included in the absorption spectrum. To include excitonic effects, the quantum-electrodynamical Bethe-Salpeter (QED-BSE) formalism \cite{Simone_nanoletter} paired with QEDFT needs to be developed for accounting for electron-hole interaction in excited setups.

The Berry curvature dipole for 2D crystalline materials is calculated using \cite{bcd_cal,PRL_Dongbin_berrydipole} 
$$D_{a} = \int d\mathbf{k} \sum_m f_m \frac{\partial \Omega_{z,m} }{\partial k_{a}},$$ 
where $\Omega_{z,m}$ is the Berry curvature and $f_m$ is Fermi-Dirac distribution for the band $m$, and $a$ represents spatial Cartesian coordinates $x, y~\text{and}~z$. The $\Omega_{z,m} $ is calculated from the tight-binding Hamiltonian, expressed in the basis set of Wannier functions using the code of Wannier90 \cite{Wannier90}, by fitting the electronic bands obtained from QEDFT. Moreover, based on the Wannierized Hamiltonian, we calculate the spectrum  
$$\eta_a = \frac{e^3\pi}{2\hbar^2} \int \frac{d \mathbf{k}}{8 \pi^3} \sum_{mn} \Delta^a_{mn} (f_n - f_m) (r_{mn}^z r_{nm}^y - r_{mn}^y r_{nm}^z) \delta(\omega_{mn} - \omega_p)$$ 
of injection current along $a$ direction \cite{injection_ACSNano,PhysRevB.61.5337}, as a function of the frequency $\omega_p$ of the applied probe light field with polarizations on $xy$ plane. $r_{nm}^a = i \langle n | \partial_{k_a} | m \rangle $ is the interband Berry connection, and $\hbar \Delta_{mn}^a$ and $\hbar \omega_{mn}$ are the difference in group velocity and the difference in energy eigenvalue between bands $n$ and $m$, respectively. The electric dipole along out-of-plane direction of 2D materials is calculated from Quantum Espresso package. 

\section{Results and discussion}
\begin{figure*}[!b]
\centering
\makebox[\textwidth]{
\includegraphics[width=1.05\textwidth]{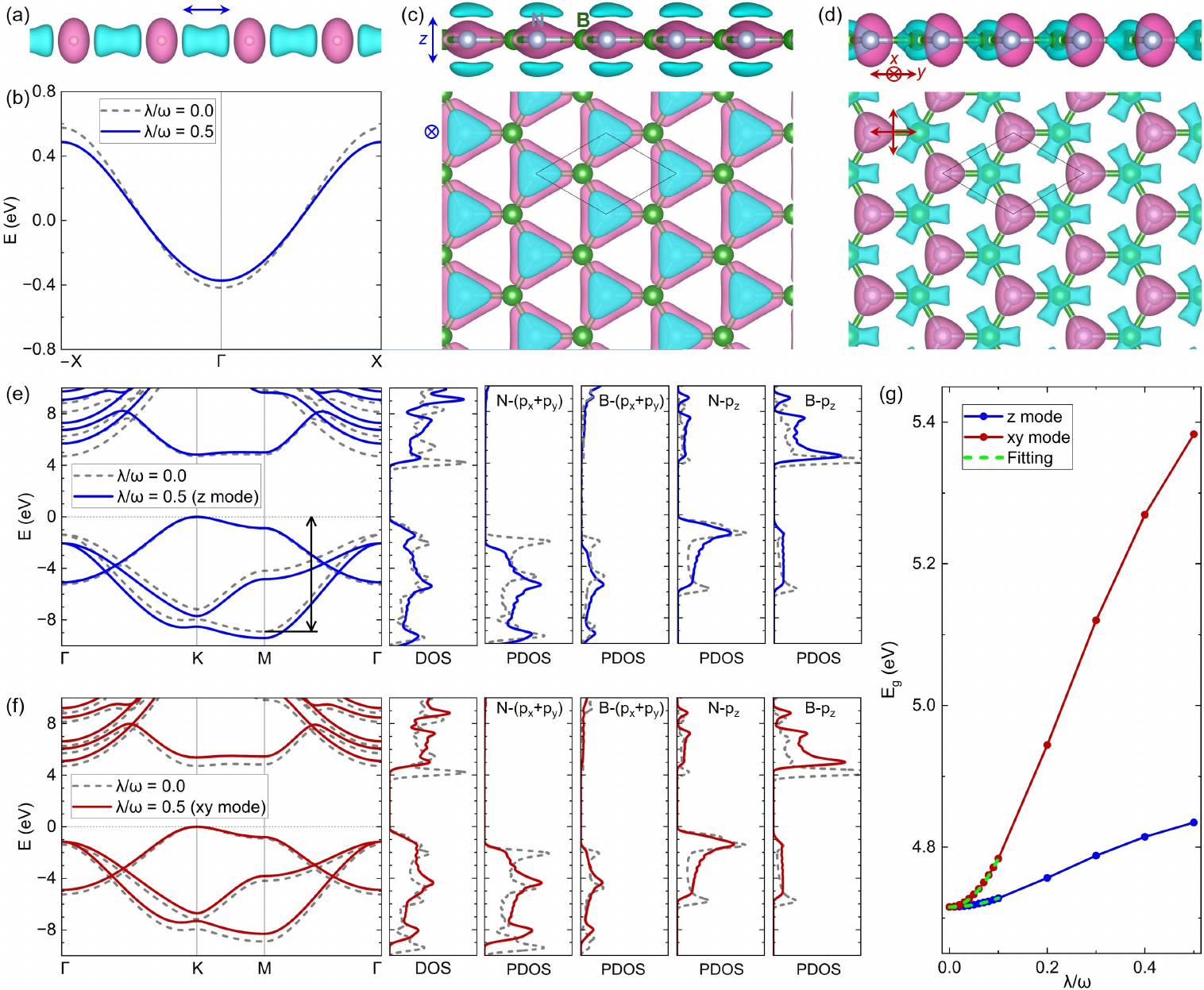}}
\caption{
Hydrogen chain and monolayer BN coupled to linearly polarized cavity photon modes. The coupling strength is $\frac{\lambda}{\omega} = 0.5$ in panels (a)-(f) for the case with light-matter coupling.
(a) Cavity-induced variation of charge density in hydrogen chain coupled to a photon polarized along chain direction.  
(b) Electronic bands of the hydrogen chain. 
(c,d) Cavity-induced variation of the electron density in monolayer h-BN coupled to (c) a photon polarized along $z$ direction, i.e., perpendicularly to h-BN on $xy$ plane and (d) two photons polarized along $x$ and $y$ directions. 
(e,f) Electronic bands, DOS, and PDOS of h-BN coupled with (e) the $z$ photon mode and (f) the $x$ and $y$ photon modes. The bandwidth of valence bands is indicated by the line with arrows on both ends.
(g) Evolution of the energy band gap of monolayer h-BN with coupling strength $\frac{\lambda}{\omega}$. The green dashed lines show the numerical fitting in the regime $0 \leq \frac{\lambda}{\omega} \leq 0.1$.
The isosurfaces in (a), (c), and (d) correspond to values of $\Delta \rho = \pm 0.0004$, $\pm 0.002$, and $\pm 0.004$ $e/\text{Bohr}^3$, respectively. The purple isosurfaces represent electron gain, while the cyan isosurfaces represent electron loss. 
}
\label{fig:hydrogen_BN}
\end{figure*}

To illustrate the effect of cavity vacuum photon fields on the electronic states of a crystalline material, we first look at an ideal 1D hydrogen chain. 
As shown in Fig.~\ref{fig:hydrogen_BN}(a,b), the hydrogen chain with an inter-site distance $3.2$~\AA~ (assuming clamped ions) exhibits metallic band dispersion with an energy width $0.994$~eV when outside an optical cavity. 
Upon interacting with a photon mode that has linear polarization along the chain ($x$) direction and coupling strength $\frac{\lambda}{\omega}=0.5~(0.1)$, the band width decreases to $0.861$~eV ($0.987$~eV). The coupling strength around $\frac{\lambda}{\omega}=0.1 $ can be achieved using phonon- and plasmon-polariton based cavities \cite{I-Te_PNAS,shin2025,boström2025,RN9848,RN9849}.
To understand the band renormalization, we evaluate the cavity-induced charge density change, $\Delta \rho = \rho(\frac{\lambda}{\omega} = 0.5) - \rho(\frac{\lambda}{\omega} = 0)$. Figure \ref{fig:hydrogen_BN}(a) shows charge accumulation (purple isosurface) at the hydrogen atoms and charge depletion (cyan isosurface) in the intermediate regions, indicating the localization of charge density---which is analogous to dynamical charge localization \cite{Simone_PNAS,dunlap1986dynamic}---at hydrogen atoms in real space, along the polarization direction of photon mode. This charge localization reduces the overlap of electron clouds between neighboring hydrogen atoms along the $x$ direction, thereby narrowing the band dispersion along the $k_x$ direction in reciprocal space [Fig. \ref{fig:hydrogen_BN}(b)]. 
The effect of charge localization along photon polarization directions is general, and it supports our previous QEDFT study on the cavity-enhanced 
electron-phonon coupling in MgB$_2$ \cite{I-Te_PNAS} and analytical study on cavity-induced para-ferroelectric transition in SrTiO$_3$ \cite{Simone_PNAS}.

Next, we look at 2D materials, starting with a widely studied band gap insulator, h-BN. 
As shown in Fig.~\ref{fig:hydrogen_BN}(c,e), a monolayer of h-BN on the $xy$ plane is insulating with a direct energy band gap. 
The coupling of h-BN to a cavity photon, with $\frac{\lambda}{\omega}=0.5$ and linear polarization along out-of-plane ($z$) direction, localizes the charge density along $z$ direction. This leads to charge accumulation in the $xy$ plane, enhancing the in-plane overlap of electron clouds between atoms. 
As a result, the bandwidth of valence bands in Fig.~\ref{fig:hydrogen_BN}(e) increases from $8.90$ eV to $9.41$ eV, corresponding to a downward energy shift of the occupied $p_{x}$ and $p_y$ orbitals of B and N atoms, as shown in the total density of states (DOS) and projected DOS (PDOS) in Fig.~\ref{fig:hydrogen_BN}(e). 
In addition, since the band edges are determined by the $p_z$ orbitals of B and N atoms, the upward energy shift of the unoccupied B-$p_z$ and N-$p_z$ orbitals leads to an increase of the direct band gap. 

Differently, the coupling of h-BN to two cavity photons, with in-plane $x$ and $y$ linear polarizations and $\frac{\lambda}{\omega}=0.5$, narrows the bandwidth of the bottom bands from $8.90$ eV to $8.30$ eV, as shown in Fig.~\ref{fig:hydrogen_BN}(d,f). 
This originates from charge accumulation at the N atoms and charge depletion at the B atoms along the in-plane directions. 
The reduction of the band dispersion is primarily due to the decreased energy spread of the occupied $p_x$ and $p_y$ orbitals of B and N atoms, as shown by the PDOS in Fig.~\ref{fig:hydrogen_BN}(f). 
Besides, the upward energy shift of the unoccupied $p_z$ orbitals of B and N atoms leads to an increase of the direct band gap. 
The tunability of direct band gap is further examined in relation to the strength of light-matter coupling and the orientation of photon polarization. 
As shown in Fig.~\ref{fig:hydrogen_BN}(g), the two cavity photons with in-plane $x$ and $y$ polarizations significantly alter the band gap compared to a cavity photon with out-of-plane ($z$) polarization, where the gap variation $\Delta E_g = E_g (\frac{\lambda}{\omega}=0.5) - E_g (\frac{\lambda}{\omega}=0)$ shows $\Delta E_g^{xy} \sim 3\Delta E_g^z$.
In the regime $0 \leq \frac{\lambda}{\omega} \leq 0.1$, the numerical fitting in Fig.~\ref{fig:hydrogen_BN}(g) shows the band gap scales quadratically with coupling strength by $E_g(\frac{\lambda}{\omega}) = E_g(\frac{\lambda}{\omega}=0) + \kappa (\frac{\lambda}{\omega})^2$, where the  coefficient is $\kappa^{z} = 6.85$ eV and $\kappa^{xy} = 1.25$ eV for $z$ and $xy$ modes, respectively. This quadratic dependence originates from the fact that, in the low-coupling limit, the photon-induced electron interaction and its corresponding electron-photon exchange functional are proportional to the square of the coupling strength $\frac{\lambda}{\omega}$ \cite{QED_functional_Christian,I-Te_PRA_QEDFT}, with a prefactor that depends on electron momentum and photon polarization.
A similar quadratic dependence is also expected in the modification of the charge density, consistent with the quadratic scaling of the electron-photon exchange potential in the low-coupling regime. The charge localization and its dependence on coupling strength might be observed through real-space probes, like electron microscopy.
Thus, the cavity-induced charge localization in monolayer h-BN is highly sensitive to the polarization directions of the cavity photon modes, and the resulting different renormalizations of electronic bands exhibit the orbital-dependent engineering for the Bloch states. 

\begin{figure*}[!t]
\centering
\makebox[\textwidth]{
\includegraphics[width=0.685\textwidth]{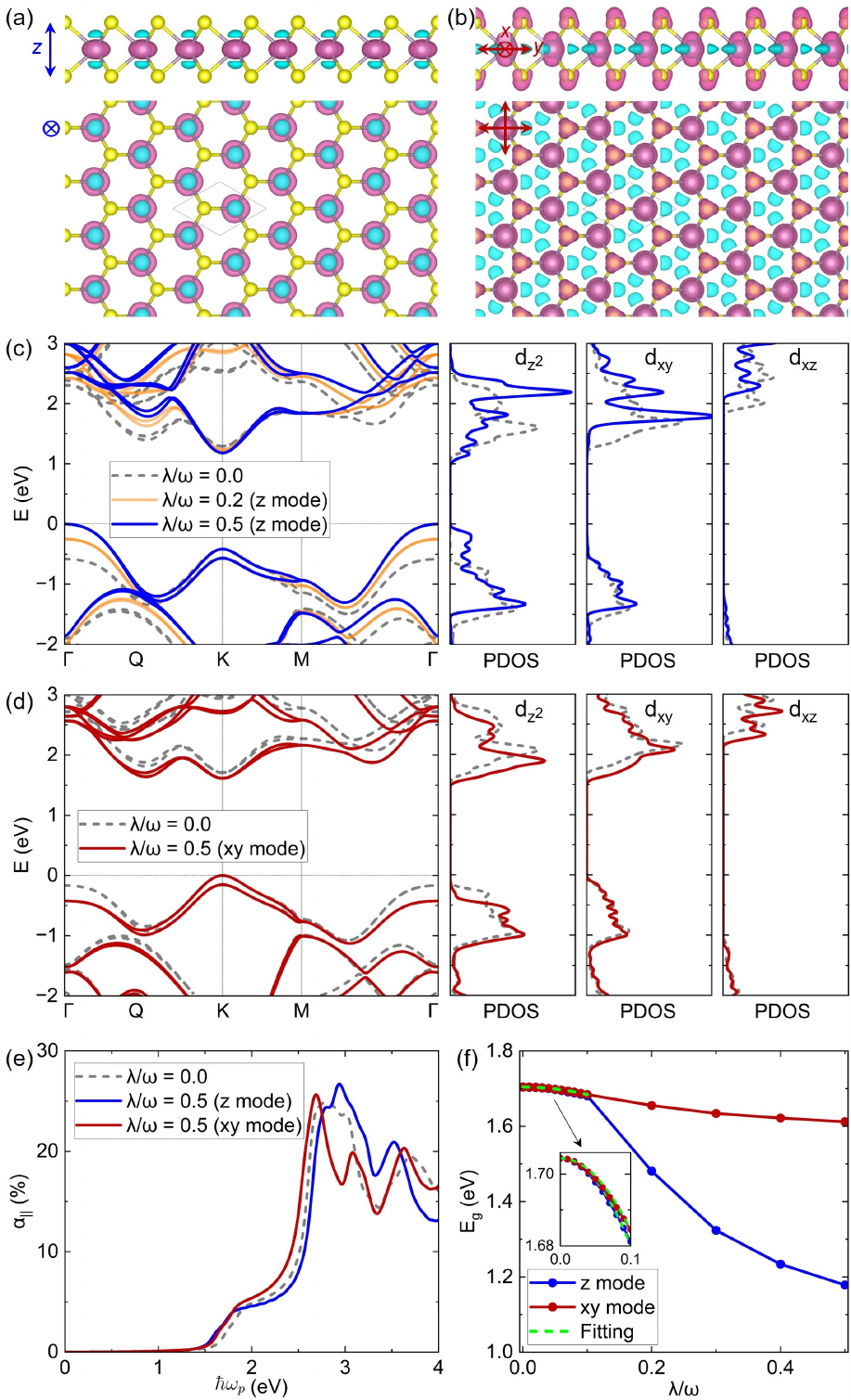}}
\caption{
Monolayer 2H-MoS$_2$ coupled to linearly polarized cavity photons. The coupling strength is $\frac{\lambda}{\omega} = 0.5$  in panels (a)-(e) for the case with light-matter coupling. 
(a,b) Cavity-induced variation of the electron density in 2H-MoS$_2$ coupled to (a) a photon polarized along out-of-plane ($z$) direction and (b) two photons polarized along in-plane ($x$ and $y$) directions. 
(c,d) Electronic bands and PDOS of 2H-MoS$_2$ coupled to (c) the $z$ photon mode and (d) the $xy$ photon modes. The PDOS for $d$ orbitals of Mo atoms are plotted, where the ${d}_{xy}$ and ${d}_{{x}^2-{y}^2}$ (also ${d}_{xz}$ and ${d}_{yz}$) are degenerate. The bands for the $z$ mode with $\frac{\lambda}{\omega} = 0.2$ is also shown in panel (c).
(e) Absorption spectrum of the 2H-MoS$_2$ coupled with the $z$ photon mode or the $x$ and $y$ modes. Electron-hole interaction is not included.
(f) Evolution of the energy band gap with coupling strength $\frac{\lambda}{\omega}$. The green dashed lines show the numerical fitting in the regime $0 \leq \frac{\lambda}{\omega} \leq 0.1$.
The isosurfaces in (a) and (b) are for the value of $\Delta \rho = \pm 0.0017$ $e/\text{Bohr}^3$, where the purple and cyan color represents electron gain and loss, respectively.  
}
\label{fig:monolayer_MoS2}
\end{figure*}

Since transition-metal dichalcogenides (TMDs) have gained a large attention for light-matter interactions \cite{RevModPhys.90.021001,Manzeli2017}, we investigate the effects of cavity-induced  charge localization on the electronic, atomic, and optical properties of monolayer and bilayer TMDs.
As shown in Fig.~\ref{fig:monolayer_MoS2}(a), for the monolayer 2H-MoS$_2$ coupled to a photon with $\frac{\lambda}{\omega}=0.5$ and linear polarization along out-of-plane ($z$) direction, the charges move along $z$ direction to accumulate at the Mo atoms.
This raises the top valence band with Mo-$d_{z^2}$ orbital composition at the crystal momentum $\Gamma$ by 0.58 eV, hence transforming the direct band gap of 1.71 eV of bare monolayer 2H-MoS$_2$ into an indirect band gap of 1.18 eV with the valence and conduction band edges at crystal momenta $\Gamma$ and K, respectively, as shown in Fig.~\ref{fig:monolayer_MoS2}(c). 
Notably, the critical point for the direct-indirect gap transition occurs at $\frac{\lambda}{\omega}=0.11$, a value that is experimentally attainable \cite{I-Te_PNAS,RN9848,RN9849}
, where the top valence band at $\Gamma$ is raised by 0.17 eV. Below this threshold, the band gap remains direct. For instance, with $\frac{\lambda}{\omega}=0.1$, a photon shift the top valence band at $\Gamma$ upward by 0.13 eV---insufficient to surpass the valence band maximum at K.
As shown in Fig.~\ref{fig:monolayer_MoS2}(c), when monolayer 2H-MoS$_2$ couples to a $z$-polarized mode with $\frac{\lambda}{\omega} = 0.2$, the top valence band at the $\Gamma$ point shifts upward by a noticeable amount 0.33 eV, which is qualitatively similar to the shift observed for $\frac{\lambda}{\omega} = 0.5$.
Figure~\ref{fig:monolayer_MoS2}(e) shows the optical absorption spectrum for a probing filed with in-plane linear polarization, and indicates that the narrowing of band gap by fluctuating cavity photons can be optically measured. 
The critical strength $\frac{\lambda}{\omega}$ for phase transition is expected to decrease for the monolayer 2H-TMDs with a smaller energy difference at $\Gamma$ and K of the top valence band.

In contrast, with two in-plane polarized cavity photons, the charges of all atoms localize along the in-plane directions, as shown in Fig.~\ref{fig:monolayer_MoS2}(b).
This shifts the energy of the top valence band at crystal momentum $\Gamma$ downward, and the band gap remains direct [Fig. \ref{fig:monolayer_MoS2}(d)]. 
Meanwhile, the conduction band minimum (CBM), contributed by Mo-$d_{z^2}$ and Mo-$d_{xy}$ orbitals, is shifted downward, giving rise to the decreasing of the direct band gap. 
The renormalization of the band gap by cavity photons can be observed by measuring the optical absorption spectrum in Fig.~\ref{fig:monolayer_MoS2}(e). Electron-hole interaction can be further included in the absorption spectrum by combining QEDFT with QED-BSE, where the inclusion of excitonic effects might shift or split the absorption peaks shown here.
The modification of band gap is further compared with respect to the strength of light-matter coupling and the polarization directions of the cavity photon modes. 
As shown in Fig.~\ref{fig:monolayer_MoS2}(f), the cavity photon with out-of-plane ($z$) polarization significantly decreases the band gap by 0.53 eV, 
compared to the gap reduction of 0.1 eV 
by two in-plane polarized cavity photons. Importantly, the transition between direct and indirect band gap is only induced by the cavity photon with $z$ polarization. 
These results indicate that an optical cavity can modulate the band gap and optical response of monolayer 2H- MoS$_2$ through charge localization, which strongly depends on the polarization directions of cavity photon modes. 
In addition, in the regime $0 \leq \frac{\lambda}{\omega} \leq 0.1$, numerical fitting reveals that the band gap scales quadratically with coupling strength by $E_g(\frac{\lambda}{\omega}) = E_g(\frac{\lambda}{\omega}=0) + \kappa (\frac{\lambda}{\omega})^2$, with coefficient $\kappa^{z} = -2.43$ eV and $\kappa^{xy} = -2.00$ eV for $z$ and $xy$ modes, respectively.

\begin{figure}[!t]
\centering
\includegraphics[width=0.75\textwidth]{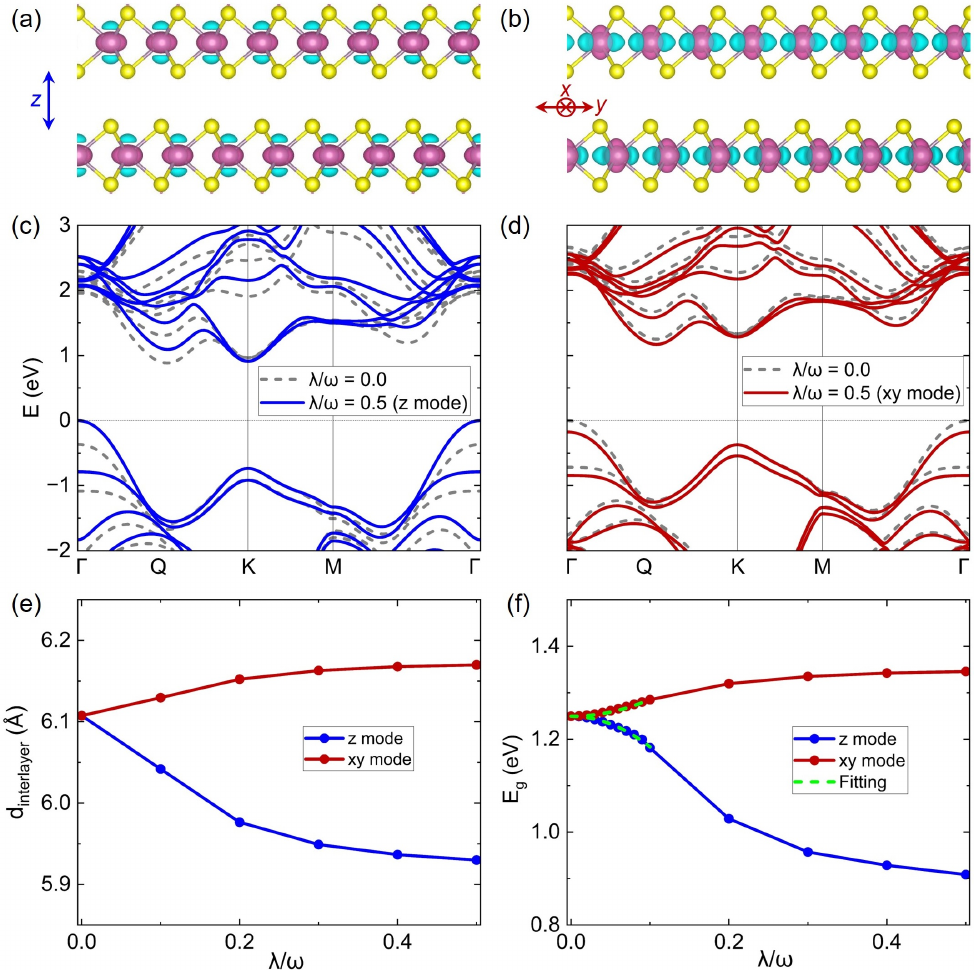}
\caption{
Bilayer 2H-MoS$_2$ coupled to linearly polarized cavity photons. The coupling strength is $\frac{\lambda}{\omega} = 0.5$  in panels (a)-(d) for the case with light-matter coupling.
(a,b) Cavity-induced variation of electron density in the bilayer 2H-MoS$_2$ coupled to (a) a photon polarized along out-of-plane $z$ direction and (b) two photons polarized along in-plane $x$ and $y$ directions. 
(c) Electronic bands modified by the mode with $z$ polarization. 
(d) Electronic bands modified by the two modes with $x$ and $y$ polarizations. 
(e) Evolution of interlayer spacing (measured as vertical distance of Mo atoms in the two layers) with the $\frac{\lambda}{\omega}$.
(f) Evolution of energy band gap with the $\frac{\lambda}{\omega}$. The green dashed lines show the numerical fitting in the regime $0 \leq \frac{\lambda}{\omega} \leq 0.1$.  
The isosurfaces in (a) and (b) are for the value of $\Delta \rho = \pm 0.00085$ $e/\text{Bohr}^3$, where the purple and cyan colors represent electron gain and loss, respectively.
}
\label{fig:bilayer_MoS2}
\end{figure}

To show the influence of fluctuating cavity photons on the interlayer interactions in 2D vdW layered materials, we investigate how cavity-induced charge localization modifies the properties of bilayer 2H-MoS$_2$.
As shown in Fig.~\ref{fig:bilayer_MoS2}(a), by coupling to a photon mode polarized in the out-of-plane ($z$) direction, the charge in each layer accumulates at Mo atoms along $z$ direction, similarly to what observed for the single layer in Fig.~\ref{fig:monolayer_MoS2}(a).
This gives rise to a decrease of 0.34 eV in the electronic band gap due to the upward energy shift of valence band maximum (VBM) at crystal momentum $\Gamma$ [Fig.~\ref{fig:bilayer_MoS2}(c)]. 
On the other hand, as shown in Fig.~\ref{fig:bilayer_MoS2}(b), with two photons linearly polarized on the $xy$ plane, the charge in each layer accumulates at Mo atoms along the in-plane direction [also observed for a single layer in Fig.~\ref{fig:monolayer_MoS2}(d)], which gives rise to an increase of 0.10 eV in the electronic band gap due to the downward energy shift of VBM at crystal momentum $\Gamma$ [Fig.~\ref{fig:bilayer_MoS2}(d)]. 
Differently from the single layer, the modification of electronic states is affected by the cavity modulation of interlayer spacing.
As shown in Fig.~\ref{fig:bilayer_MoS2}(e,f), with an out-of-plane ($z$) photon mode, the decreased interlayer spacing, measured as the vertical distance between Mo atoms in the two layers,  introduces stronger hybridization between layers, contributing to a decrease in band gap.
In contrast, the increase of the interlayer spacing, predicted with in-plane ($xy$) photon modes, increases the band gap.
In the regime $0 \leq \frac{\lambda}{\omega} \leq 0.1$, the coefficient of the quadratic evolution of band gap $E_g$ with coupling strength $\frac{\lambda}{\omega}$ is $\kappa^{z} = -6.60$ eV and $\kappa^{xy} = 3.60$ eV for $z$ and $xy$ modes, respectively, where the opposite signs show gap decreasing and increasing for corresponding photon polarizations.
These results indicate that the interlayer spacing in a layered material stacked by vdW interaction can be increased or decreased by an optical cavity as a function of light-matter coupling strength and photon polarization directions, thereby contributing to non-perturbative changes of the material properties. 
Since vdW interactions are influenced by the surrounding electromagnetic vacuum, the structured vacuum fluctuations within a cavity are expected to modify vdW interactions. This modification is evidenced by the modulation of interlayer spacing in the bilayer MoS$_2$ held together by vdW dispersion forces.
We note that the cavity-modified atomic and electronic structures, calculated here for layered materials embedded in optical cavities, share similarities with the ones observed for the same materials under pressure \cite{TMD_pressure1,TMD_pressure2}. This suggests that the materials whose properties are sensitive to pressure can be also affected by cavity vacuum fluctuations.

\begin{figure}[!b]
\centering
\includegraphics[width=0.75\textwidth]{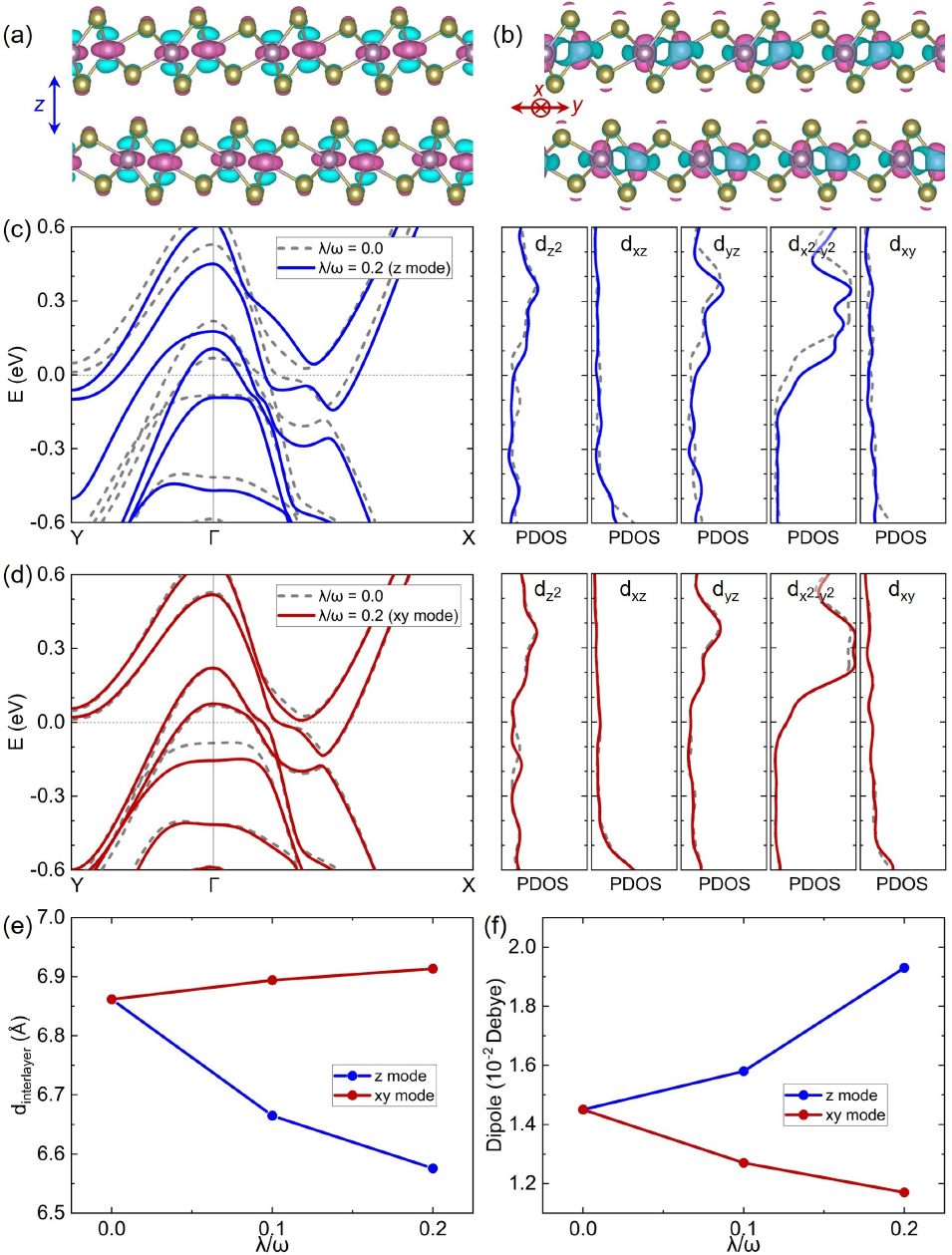}
\caption{
Bilayer T$_\text{d}$-MoTe$_2$ coupled to linearly polarized cavity photons. The coupling strength is $\frac{\lambda}{\omega} = 0.2$  in panels (a)-(d) for the case with light-matter coupling.
(a,b) Cavity-induced variation of electron density in the bilayer T$_\text{d}$-MoTe$_2$ coupled to (a) a photon polarized along out-of-plane $z$ direction and (b) two photons polarized along in-plane $x$ and $y$ directions. 
(c,d) Electronic bands and the PDOS of Mo-$d$ orbitals, modified by (c) the mode with $z$ polarization and (d) the two modes with $x$ and $y$ polarizations. 
(e) Evolution of interlayer spacing (characterized by the vertical distance of Mo atoms in the two layers) with the $\frac{\lambda}{\omega}$.
(f) Evolution of vertical electric dipole with the $\frac{\lambda}{\omega}$.  
The isosurfaces in (a) and (b) are for the value of $\pm 0.0002$ $e/\text{Bohr}^3$, where the purple and cyan colors represent electron gain and loss, respectively.
}
\label{fig4:bilayer_MoTe2}
\end{figure}

Since the out-of-plane ferroelectricity of bilayer T$_\text{d}$-MoTe$_2$ depends on the interlayer spacing, we investigate the influence of cavity photon modes on the material properties of semimetallic bilayer T$_\text{d}$-MoTe$_2$. This material exhibits an out-of-plane electric dipole polarization due to the absence of inversion symmetry. 
As shown in Fig.~\ref{fig4:bilayer_MoTe2}(a), with light-matter coupling strength $\frac{\lambda}{\omega} = 0.2$ and out-of-plane ($z$) polarization, the cavity photon localizes the charge at Mo atoms along the polarization direction. 
Similarly to that of bilayer 2H-MoS$_2$ in Fig.~\ref{fig:bilayer_MoS2}(a), the charge modification $\Delta \rho = \rho(\frac{\lambda}{\omega} = 0.2) - \rho(\frac{\lambda}{\omega} = 0)$ effectively makes each layer thinner, decreasing interlayer spacing.
Meanwhile, the electronic bands and PDOS are renormalized as in Fig.~\ref{fig4:bilayer_MoTe2}(c), where the most significant changes come from Mo-$d$ atomic orbitals.
In contrast, as shown in Fig.~\ref{fig4:bilayer_MoTe2}(b,d), with $\frac{\lambda}{\omega} = 0.2$ and in-plane polarization, the two cavity photons change the electronic structure by localizing charge of Mo-$d$ orbitals at the Mo atoms along the in-plane directions, leading to an increase of interlayer spacing. 
The decrease and increase of interlayer spacing from out-of-plane and in-plane polarized cavity photons leads to an increase and decrease in the ferroelectric dipole due to stronger and weaker charge transfer between layers, respectively, as shown in Fig.~\ref{fig4:bilayer_MoTe2}(e,f). Also, the modification of ferroelectric dipole depends on light-matter coupling strength. Thus, the ferroelectric dipole of bilayer T$_\text{d}$-MoTe$_2$ can be quantitatively controlled by changing  both photon polarization and light-matter coupling strength.

\begin{figure}[!b]
\centering
\includegraphics[width=0.85\textwidth]{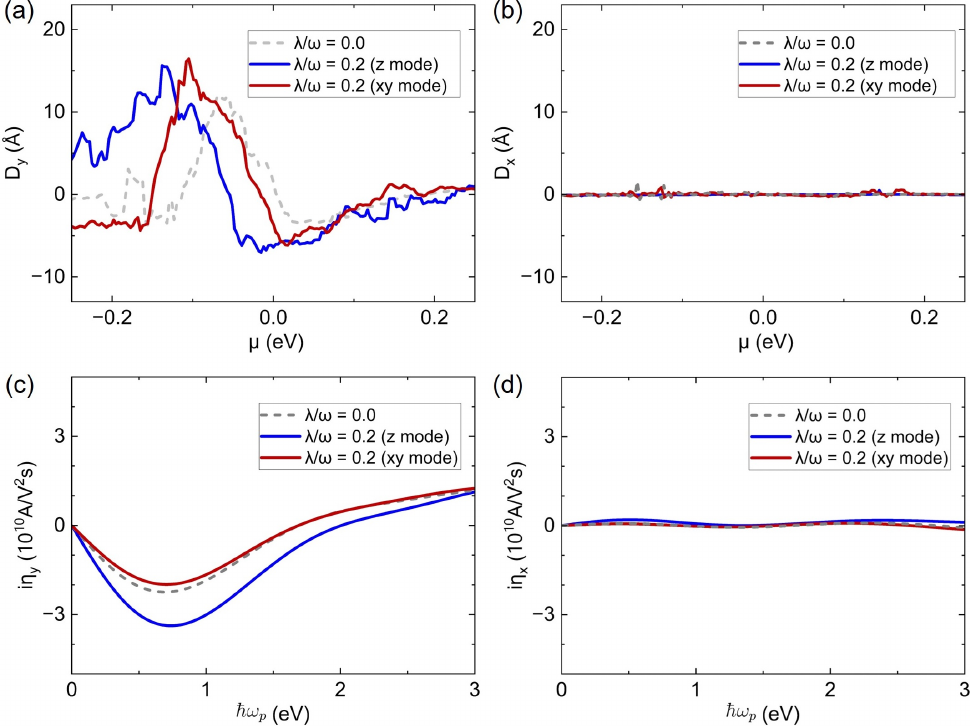}
\caption{
Cavity-modified (a,b) Berry curvature dipole $D_y$ and $D_x$ and (c,d) injection current response function $\eta_y$ and $\eta_x$ of bilayer T$_\text{d}$-MoTe$_2$ coupled to linearly polarized cavity photon modes. The coupling strength is $\frac{\lambda}{\omega} = 0.2$ for the case with nonzero light-matter coupling. 
Spin-orbit coupling is included. 
}
\label{fig:berry_injection}
\end{figure}

Since quantum geometries and optical response functions are determined by electronic structure and lattice symmetry \cite{PRL_Dongbin_berrydipole,injection_ACSNano,graphene_cavity_topology_single_particle}, the modifications of the electronic structure [Fig.~\ref{fig4:bilayer_MoTe2}(c,d)] in bilayer T$_\text{d}$-MoTe$_2$ by cavity photons manifest possible engineering on nonlinear Hall effect and circular photogalvanic effect.
As shown in Fig.~\ref{fig:berry_injection}(a,b), the Berry curvature dipole with components $D_y$ and $D_x$ is evaluated with respect to chemical potential ($\mu$) under different photon field conditions.
For the system outside the cavity, the positive peak of $D_y$ is located at $\mu = -0.06$~eV, while $D_x=0$ due to the presence of a $yz$ mirror plane in atomic structure.
When the material is coupled with to a cavity photon mode with strength $\frac{\lambda}{\omega}=0.2$ and along the out-of-plane ($z$) direction, the $D_y$ is significantly modified, with the peak shifting to $\mu = -0.14$ eV.
A similar shift to $\mu = -0.11$ eV is observed for photon modes with in-plane polarization.
This shows that fluctuating cavity photon can be exploited to engineer the nonlinear anomalous Hall effect.
Moreover, as a representative example of second-order optical response, the injection current  $\eta_y$ can also be controlled via coupling to cavity photons, as shown in Fig.~\ref{fig:berry_injection}(c,d), while the current $\eta_x = 0$ due to the presence of a $yz$ mirror plane. 
For $\frac{\lambda}{\omega}=0.2$, the coupling with a cavity photon with out-of-plane ($z$) polarization enhances the peaked profile of $\eta_y$ around 0.70 eV by 50.5\%, while the photons with in-plane $x$ and $y$ polarizations weaken the peak by 11.5\%. 
Thus, the injection current can be controlled by changing the polarization of photon modes. 
These results show that strong light-matter interaction via vacuum-fluctuating photon fields can also modify the transport and optical response properties of 2D layered TMDs embedded in optical cavities or other electromagnetic environments. The tunability of the Berry curvature and injection current roots in the renormalization of electron momentum by fluctuating photons, described by the minimal electron-photon coupling $\hat{\mathbf{p}}_i - \hat{\mathbf{A}}$, which indicates quantum geometry of Bloch electronic wavefunctions can be modified by cavity photons.

In this study, the quantitative predictions from QEDFT are presented for a single $z$ mode and two equivalent $xy$ modes. The inclusion of more photon modes is expected to yield qualitatively similar modifications, with differences appearing primarily in the quantitative details. Notably, the critical coupling strength for phase transitions may decrease due to the cumulative effect of multiple photon modes. 
Also, unlike the $x$ and $y$ modes with the same coupling strength in this study, two or more in-plane-polarized modes with different coupling strength could be employed to alter the spatial symmetries of crystals, potentially enabling symmetry-breaking phenomena. The change of spatial symmetries by multiple anisotropic photon modes may be observed by measuring the photon-induced variation of charge density in real space.

We estimate the regime of $\frac{\lambda}{\omega}$ that can be achieved at present. In experiments, mode strength is usually characterized by electric field amplitude $\mathcal{E}$ of photon field. Based on the relation $\lambda = \frac{\sqrt{2} \mathcal{E}}{\sqrt{\omega}}$ (in atomic units), the achievable coupling strength $\frac{\lambda}{\omega}$ falls in the range  of approximately 0.03 to 0.3, as inferred from the experimentally reported value of $\mathcal{E}$ for phonon- and plasmon-polariton based cavities \cite{I-Te_PNAS,RN9849,RN9848}.  While the strong coupling of $\frac{\lambda}{\omega} = 0.5$ is difficult to be reached in experiments at present, the photon-induced phenomena, including the band gap modification from energy shift, can be observed in the experimentally accessible coupling strengths up to $\frac{\lambda}{\omega}=0.3$. As shown in Figs. \ref{fig:hydrogen_BN}(g), \ref{fig:monolayer_MoS2}(f) and \ref{fig:bilayer_MoS2}(f), the band gap is modified by a noticeable quantity at the more realistic coupling strengths. Specifically, for a $z$ mode with $\frac{\lambda}{\omega}=0.2$, the change $\Delta E_g$ in band gap is 0.04 eV for monolayer BN, and 0.22 eV for monolayer and bilayer 2H-MoS$_2$. This indicates that the QEDFT-based predictions presented in this work may be experimentally realizable using phonon- or plasmon-polariton-based cavities, particularly within the coupling regime $\frac{\lambda}{\omega} = 0$ to 0.3.

\section{Conclusion}
In conclusion, our state-of-the-art QEDFT simulations reveal the significant effects of fluctuating cavity photon fields on the vdW quantum materials.
Cavity-induced dynamical charge localization along the photon polarization directions is generally observed in low-dimensional materials, enabling the modification of vdW dispersion forces and the flexible engineering of various atomic, electronic, optical, and transport properties.
At the 1D hydrogen chain, charge localization reduces electron kinetic energy, as evident from the suppressed band dispersion. 
For 2D vdW layered material systems, photon modes with out-of-plane and in-plane linear polarizations localize charge density in each layer along out-of-plane and in-plane directions, respectively. 
Beyond influencing the energy band width, fluctuating cavity photons also modulate the band gaps in monolayer h-BN and 2H-MoS$_2$ by altering the energies of atomic orbitals at the VBM and CBM, thereby tuning the absorption spectra. Notably, out-of-plane polarized photons induce a transition from a direct to an indirect band gap in monolayer 2H-MoS$_2$.
Moreover, photon modes with different polarizations adjust the interlayer spacing in bilayer systems: out-of-plane polarized photons reduce the spacing, while in-plane polarized photons expand it in semiconducting bilayer 2H-MoS$_2$ and semimetallic bilayer T$_\text{d}$-MoTe$_2$. This tunability enables control over the ferroelectric dipole in bilayer T$_\text{d}$-MoTe$_2$. Also, the nonlinear Hall effect and the circular photogalvanic effect in the bilayer T$_\text{d}$-MoTe$_2$ can be modified by fluctuating cavity photons.
Overall, these findings underscore the sensitivity of material properties in 2D vdW crystals to the polarization directions of fluctuating photon modes through cavity-induced dynamical charge localization. Our results highlight the potential of strong collective light-matter interactions in optical cavities to modulate and control a broad range of properties in vdW material systems. The insights pave the way for the prediction and understanding of cavity-induced phenomena in other 2D layered crystals, such as understanding of transitions between different charge density wave phases in the correlated material 1T-TaS$_2$ \cite{Jarc2023_TaS2_cavity}. Additionally, to experimentally verify the predictions in this QEDFT study, cavity setups based on interface structures or incorporating plasmonic or phonon polaritons may be suitable. When the mode strength, frequency, and polarization of for realistic cavities have been determined from cavity design principle, these modes are used as input for QEDFT to model realistic light-matter coupled systems. This integration of QEDFT with optical cavity design is a promising avenue for future research. 

\begin{backmatter}
\bmsection{Funding}
We acknowledge financial support from the Cluster of Excellence 'CUI: Advanced Imaging of Matter'- EXC 2056 - project ID 390715994, SFB-925 "Light induced dynamics and control of correlated quantum systems" - project ID 170620586  of the Deutsche Forschungsgemeinschaft (DFG), the European Research Council (ERC-2024-SyG-UnMySt - 101167294), and the Max Planck-New York City Center for Non-Equilibrium Quantum Phenomena. The Flatiron Institute is a division of the Simons Foundation. 
Dongbin Shin is supported by the National Research Foundation of Korea (NRF) grant funded by the Korea government (MSIT) (No. RS-2024-00333664 and RS-2023-00218180) and the Ministry of Science and ICT (No. 2022M3H4A1A04074153). Simone Latini is supported by the European Union Marie Sklodowska-Curie Doctoral Networks SPARKLE grant (No. 101169225), and the Max Planck partner group, established with Technical University of Denmark.
Hang Liu is supported by the Alexander von Humboldt Foundation (Humboldt Research Fellowship) and National Natural Science Foundation of China (No. 12204338). This work was supported by the Max Planck Computing and Data Facility (MPCDF). 

\bmsection{Disclosures}
The authors declare no conflicts of interest.

\bmsection{Data Availability Statement}
Data underlying the results presented in this paper are not publicly available at this time but may be obtained from the authors upon reasonable request.
 
\end{backmatter}

\bibliography{main}

\end{document}